\documentclass[twoside,slac_one]{revtex4}
\usepackage{graphicx}
\usepackage{fancyhdr}
\usepackage{amsmath} 
\usepackage{bm}
\usepackage{amsxtra}
\usepackage{amssymb}
\usepackage{amsthm}
\usepackage{latexsym}
\usepackage{lscape}

\begin{document}

\title{Measurement of the Cross Section for Prompt Isolated Diphoton
       Production in $p\overline{p}$ Collisions at $\sqrt{s}=1.96$ TeV}

%

\author{Costas Vellidis for the CDF Collaboration}
\affiliation{Fermi National Accelerator Laboratory, Batavia, IL, USA}

\begin{abstract}
This article reports a measurement of the cross section of prompt isolated
photon pair production in $p\overline{p}$ collisions at a total energy
$\sqrt{s}=1.96$ TeV using data of 5.36 fb$^{-1}$ integrated luminosity
collected with the CDF II detector at the Fermilab Tevatron. The measured
cross section, differential in basic kinematic variables, is compared with
three perturbative QCD predictions, a leading order (LO) parton shower
calculation and two next-to-leading order (NLO) calculations. The NLO
calculations reproduce most aspects of the data. By including photon
radiation from quarks before and after hard scattering, the parton shower
prediction becomes competitive with the NLO predictions.
\end{abstract}

\maketitle

\thispagestyle{fancy}


\section{Introduction}
The measurement of the production cross section of two energetic isolated
central photons (diphotons) in high energy hadron collisions is important
for testing standard model (SM) predictions in the domain of searches for
undiscovered particles and new physics. Photons originating from hard
collisions of hadrons (``direct'' or ``prompt'' photons) are an ideal probe
for testing perturbative Quantum ChromoDynamics (pQCD) and soft-gluon
resummation methods implemented in theoretical calculations because they do
not interact with other final state particles, and their energies and
directions can be measured with high precision in modern electromagnetic
calorimeters. Prompt diphoton production with large invariant mass creates
an irreducible background in searches for a low mass Higgs boson decaying
into a photon pair \cite{higgs}, as well as in searches for new phenomena,
such as new heavy resonances \cite{heavy}, extra spatial dimensions
\cite{dim,grav} or cascade decays of heavy new particles \cite{casc}.
Precise measurements of the diphoton production differential cross sections
for various kinematic variables and their theoretical understanding are thus
very important for these searches.

\begin{figure}[h]
\centering
\includegraphics[width=80mm]{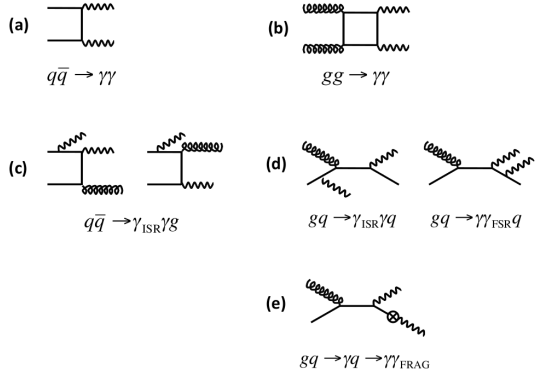}
\caption{Basic diagrams for prompt diphoton production: (a-b) direct,
     (c-d) one-photon radiation from an initial- (ISR) or final-state quark
     (FSR), (e) fragmentation where one photon is emitted along the direction
     of a final-state quark taking almost all of its energy. The symbol
     $\otimes$ denotes the non-perturbative mechanism of the fragmentation
     process (FRAG).}
\label{fig:diagrams}
\end{figure}

The basic mechanisms of prompt diphoton production in hadron collisions are
quark-antiquark annihilation $q\bar q\rightarrow\gamma\gamma$, quark-gluon
scattering $gq\rightarrow\gamma\gamma q$, and gluon-gluon fusion
$gg\rightarrow\gamma\gamma$. The respective basic diagrams are shown
in Fig. \ref{fig:diagrams}. At the Tevatron, the dominant mechanism is
quark-antiquark annihilation. In quark-gluon scattering, most of the time at
least one of the two photons is emitted almost parallel to the scattered quark.
Contributions from this mechanism are therefore suppressed by requiring
isolated prompt photons. Each mechanism can be modeled by calculating the
respective matrix element for the specific event kinematics. Matrix element
calculations of Leading Order (LO) in the strong coupling are relatively
simple and are thus implemented in advanced parton shower Monte Carlo (MC)
event generators \cite{pythia,herwig,sherpa}, which allow for gluon and
photon radiation as well as multiple interactions in the colliding beams.
By including radiation before and after the hard scattering, parton shower
generators take into account soft gluon and photon emissions, thus resulting
in an effective resummation of all of the Leading Logarithmic (LL) terms in
the cross section to all orders of the strong and electromagnetic couplings
constants. Next-to-Leading Order (NLO) calculations \cite{diphox,g2mc,resbos}
additionally include one-loop corrections at the cost of not featuring
realistic multi-particle event representations as the LO generators do.
Recent NLO calculations include an analytical resummation of the cross
section for initial-state gluon radiation to all orders in the strong
coupling constant \cite{resbos}, reaching a higher logarithmic accuracy
than in the parton shower Monte Carlo generators. By this method, all soft
gluon emissions in the initial state are taken into account, and reliable
predictions for the low diphoton transverse momentum region are possible.
A fixed-order NLO calculation implemented by the {\sc diphox} program
\cite{diphox} also accounts for the case where a final state quark loses
almost all of its energy to the photon detected in the event [diagram (e)
of Fig. \ref{fig:diagrams}] \cite{frag}. This process is called
``fragmentation'' and, in contrast to final state photon radiation in
parton showering, it involves non-perturbative calculations. One or both
photons in the event may come from fragmentation. The case where both
photons come from fragmentation of a single quark is also possible, but
is not included in calculations, as in this case the photons are nearly
collinear and thus non-isolated most of the time.

%



Diphoton measurements have been previously conducted at fixed-target
\cite{fixtgt} and collider experiments \cite{ua1,ua2,cdfdip,d0}. The most
recent measurements \cite{cdfdip,d0} were compared with the same pQCD
calculations examined in the present work and large discrepancies were
found between the data and a LO matrix element calculation supplemented
with a parton shower model, suitable for simulation of the backgrounds in
searches of a low mass Higgs boson and of new phenomena. This work shows
that the inclusion of photons radiated from initial and final state quarks
drastically improves the comparison of the parton shower calculation with
the data.

\section{Measurement}
The reported measurement was conducted using data of total integrated
luminosity 5.36 fb$^{-1}$ collected with the Collider Detector at Fermilab
(CDF) \cite{cdf} at the Tevatron $p\overline{p}$ collider. CDF is composed
of a central spectrometer inside a 1.4 T magnetic field, surrounded by
electromagnetic and hadronic calorimeters and muon detection chambers. The
inner spectrometer measures charged particle tracks with a transverse
momentum ($p_{\rm T}$) precision of
$\Delta p_{\rm T}/p_{\rm T}^{2}=0.07\%({\rm GeV/c})^{-1}$. The central
calorimeters cover the region $\vert\eta\vert<1.1$, with an electromagnetic
(hadronic) energy resolution of
$\sigma(E_{\rm T})/E_{\rm T}=13.5\%/\sqrt{E_{\rm T}({\rm GeV})}\oplus 1.5\%$
($\sigma(E_{\rm T})/E_{\rm T}=50\%/\sqrt{E_{\rm T}({\rm GeV})}\oplus 3\%$) and a
tower segmentation of $\Delta\eta\times\Delta\phi\simeq 0.1\times 15^{\circ}$.
Photons are reconstructed in clusters of up to three towers \cite{phosel}.
$\chi^{2}$ criteria are imposed on the profile of the shower to match
expected patterns. Two main cuts are applied: (i) the photon transverse
energy is required to be $E_{\rm T}\ge 17$ GeV for the first photon in the
event and $E_{\rm T}\ge 15$ GeV for the second photon; (ii) the calorimeter
isolation energy in the isolation cone around each photon \cite{caliso} is
required to be less than 2 GeV.

The background from $\gamma$$+$jet and dijet events, where one or two
jets are faking a photon, is subtracted with a method using the track
isolation as the discriminant between signal and background \cite{trkiso}.
It is based on the substantial difference of the track isolation
distribution for signal photons (nearly exponential) and for background
photons (nearly flat). The advantages of this method are that (i) it has
little sensitivity to multiple interactions in the colliding beams, so
that the signal-background separation does not degrade at high instantaneous
luminosity, and (ii) it has high efficiency and good track momentum
resolution, implying minimal degradation of the signal-background separation
due to instrumental effects. The signal fraction is determined by summing
the probabilities of an event to be pure signal, pure background, or mixed
photon pair. These probabilities are obtained by solving a $4\times 4$ matrix
equation using the observation value (0 or 1) for all four combinations
of the leading or sub-leading photon having track isolation below or above
1 GeV/c as an input. The matrix is constructed from the $E_{\rm T}$-dependent
efficiencies of signal and background photons passing the track isolation cut.
A threshold cut of 1 GeV/c is determined by maximizing the separation between
signal and background. The efficiencies are determined from Monte Carlo (MC)
$\gamma$$+$jet and dijet samples, which are produced using the {\sc pythia}
event generator \cite{pythia}. {\sc pythia} events are fully simulated through
the detector and trigger and are reconstructed with the CDF II simulation and
reconstruction software \cite{cdfsim}. With this matrix technique the full
correlations between the two photons in the event are properly taken into
account. Tests are made for underlying event contributions in complementary
cones to the photon reconstruction cone \cite{cones} and also using isolated
tracks in dijet events. The systematic uncertainty in the signal fraction
with this method is of the order of 15-20\%.

The diphoton production cross section differential in a kinematic variable
is obtained from the histogram of the estimated signal in the selected
variable. The average cross section in a bin of the variable is determined
by dividing the bin content by the trigger efficiency, the diphoton selection
efficiency and acceptance, the integrated luminosity and the bin size. The
diphoton trigger efficiency is derived from data \cite{higgs}. It is
consistent with 100\% over all of the kinematic range with a flat
uncertainty of 3\%. The selection efficiency is determined from data and
MC with an iterative method. In the first pass the efficiency is determined
from a fully simulated and reconstructed {\sc pythia} diphoton MC sample by
dividing the number of events passing all selection cuts by the number of
events passing only the kinematic cuts on the photon $E_{\rm T}$, $\eta$,
angular separation and isolation at the event generation level. The
efficiency denominator is corrected for the ``underlying event'' from
collision remnants which make the efficiency obtained from {\sc pythia}
too high by removing events from the denominator through the isolation cut.
This correction is derived by running {\sc pythia} with and without underlying
event and amounts to a constant factor of 0.88 per event. A flat 6\%
uncertainty in the selection efficiency (3\% per photon) accounts for
possible inaccuracies in the {\sc pythia} model for the underlying event.
The signal events of the data are corrected for the preliminary efficiency.
The data are then used to reweight the {\sc pythia} events and obtain a
more accurate representation of the true diphoton distribution. The
efficiency is determined using the reweighted {\sc pythia} sample and
corrected for luminosity dependence, derived from a comparison of the
vertex multiplicity distribution in data and {\sc pythia} MC
$Z^{0}\rightarrow e^{+}e^{-}$ events. The systematic uncertainty in the
efficiency resulting from the luminosity dependent correction grows linearly
from 1.8\% for $E_{\rm T}\le 40$ GeV to 3\% at $E_{\rm T}=80$ GeV and remains
constant above this point. Finally, a 6\% constant uncertainty comes from
the Tevatron integrated luminosity \cite{lumi}.

The $Z^{0}\rightarrow e^{+}e^{-}$ sample is used for calibration by
applying a ``diphoton-like'' event selection, i.e. by imposing a diphoton
selection with the same trigger but allowing for a track associated with
each of the two electromagnetic objects in the event. The electromagnetic
energy scale in data and MC is corrected by tuning the
$Z^{0}\rightarrow e^{+}e^{-}$ mass peak to the world average \cite{zmass}
and a systematic uncertainty from this correction is estimated to grow
linearly from 0 at $E_{T}\le 40$ GeV up to 1.5\% at $E_{\rm T}=80$ GeV
and remain constant above this point. The difference in the photon
identification efficiency between data and MC is estimated from the
$Z^{0}\rightarrow e^{+}e^{-}$ sample \cite{higgs} and added as a systematic
uncertainty to the measurement. All systematic uncertainties in the cross
section measurement are added in quadrature.

\section{Predictions}
\begin{figure}[t]
  \centering
    \includegraphics[width=58mm]{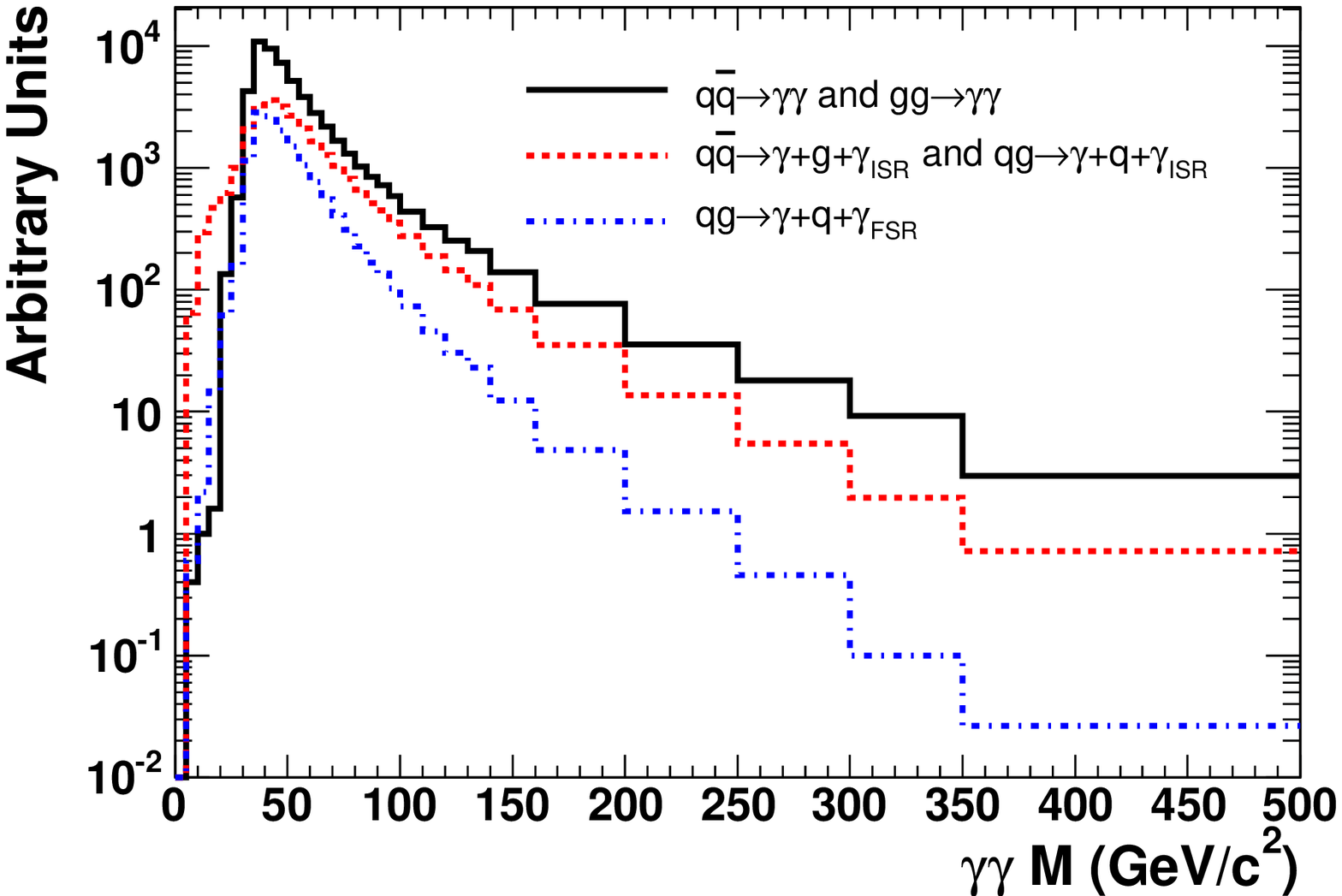}
    \hspace{-0.5cm}
    \includegraphics[width=58mm]{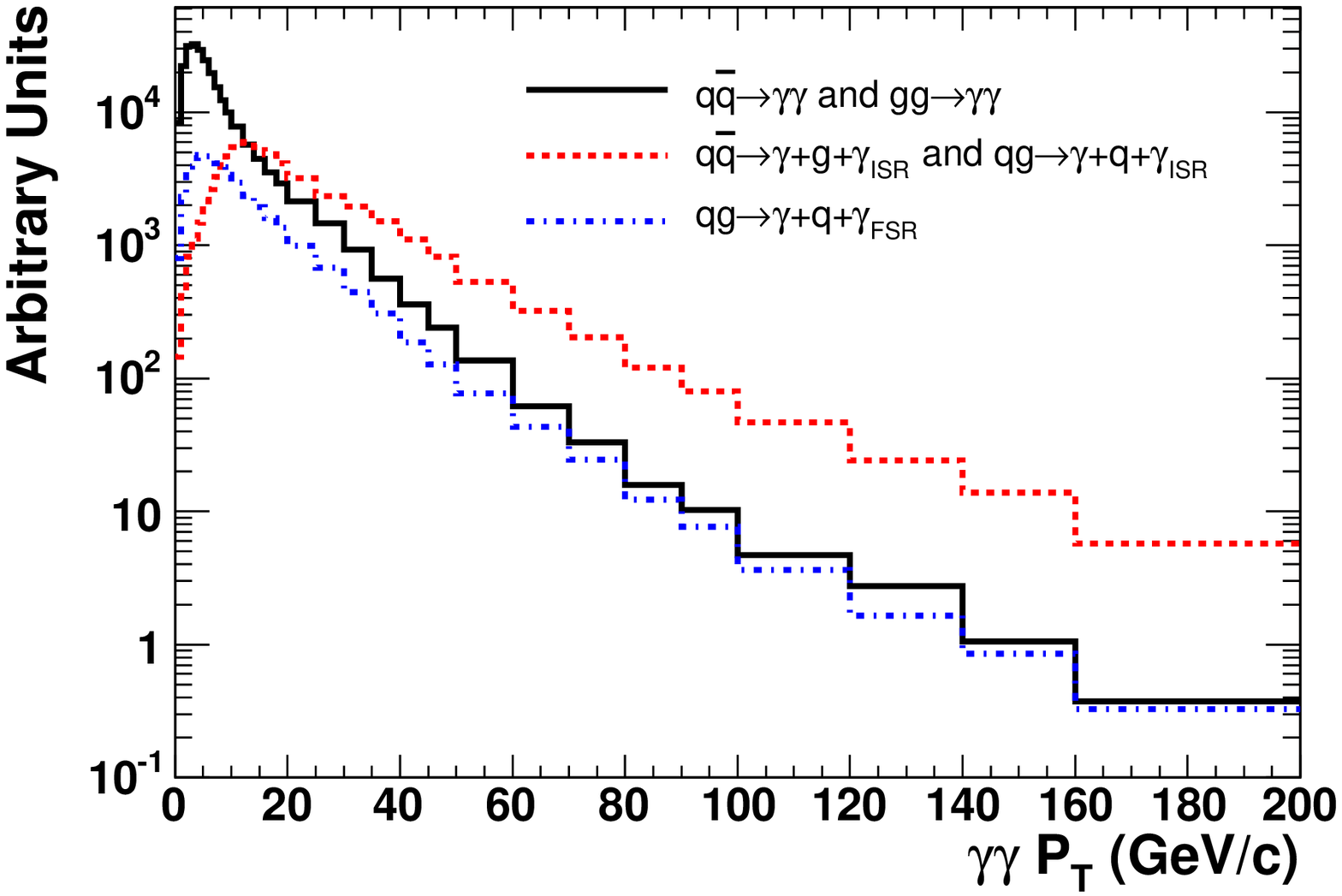}
    \hspace{-0.5cm}
    \includegraphics[width=58mm]{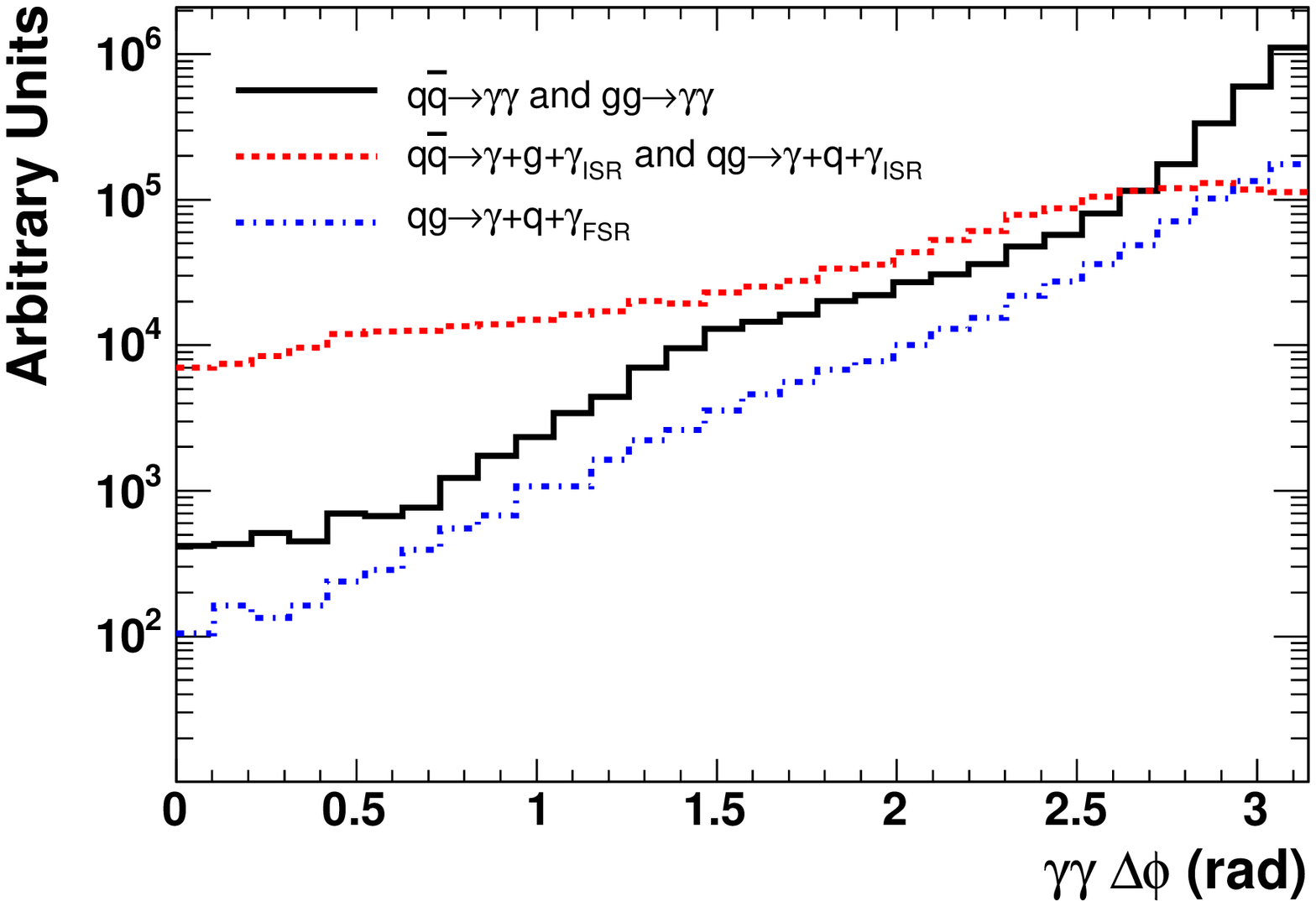}
    \caption{The individual contributions to the cross section from events
             where both photons are generated according to the {\sc pythia}
             diphoton matrix element and from events where one photon
             originates from initial or final state radiation, as functions
             of the diphoton mass (left), transverse momentum (middle) and
             azimuthal difference (right).}
\label{fig:pythia_breakdown}
\end{figure}

The results of this measurement are compared with three theoretical
calculations: (i) the fixed NLO predictions of the {\sc diphox} program
\cite{diphox} including parton fragmentations into photons \cite{frag},
(ii) the predictions of the {\sc resbos} program \cite{resbos} where the
cross section is accurate to NLO, but also has an analytical initial state
soft gluon resummation, and (iii) the predictions of the {\sc pythia} program
\cite{pythia} which features a realistic representation of the physics events
by including parton showering, Initial (ISR) and Final State Radiation (FSR)
and an underlying event model. Diphoton events were selected from an inclusive
$\gamma$$+$X {\sc pythia} sample (X$=$$\gamma$ or jet), thus including the
$q\bar q\rightarrow\gamma\gamma$ and $gg\rightarrow\gamma\gamma$ processes
(56$\%$) as well as the $q\bar q\rightarrow g\gamma\gamma_{_{\rm ISR}}$,
$gq\rightarrow q\gamma\gamma_{_{\rm ISR}}$ and
$gq\rightarrow q\gamma\gamma_{_{\rm FSR}}$ processes (44$\%$). This type of
calculation effectively resums the cross section for gluon and photon
radiation both in the initial and the final state. Fig.
\ref{fig:pythia_breakdown} shows the individual contributions to the cross
section as a function of the diphoton invariant mass, transverse momentum
and azimuthal difference. Initial-state radiation (ISR) photons, in particular,
produce substantially different distributions than ME and final-state radiation
(FSR) photons, having a harder transverse momentum spectrum and stronger
low--$\Delta\phi$ tail in the azimuthal difference spectrum. In leading order,
this can be attributed to the fact that FSR occurs in quark-gluon scattering
[diagram (d) of Fig. \ref{fig:diagrams}], whereas ISR occurs both in $q\bar q$
annihilation [diagram (c) of Fig. \ref{fig:diagrams}] and quark-gluon
scattering, and the luminosity of quark-gluon states falls off more rapidly
with the parton momenta than the luminosity of $q\bar q$ states
\cite{diphox,resbos}. Double radiation processes in minimum bias dijet events,
such as $qq\rightarrow qq\gamma_{_{\rm ISR/FSR}}\gamma_{_{\rm ISR/FSR}}$,
$q\bar q\rightarrow q\bar q\gamma_{_{\rm ISR/FSR}}\gamma_{_{\rm ISR/FSR}}$,
$gq\rightarrow gq\gamma_{_{\rm ISR/FSR}}\gamma_{_{\rm ISR/FSR}}$,
$q\bar q\rightarrow gg\gamma_{_{\rm ISR}}\gamma_{_{\rm ISR}}$ and
$gg\rightarrow q\bar q\gamma_{_{\rm FSR}}\gamma_{_{\rm FSR}}$, were also examined
but their overall contribution was estimated to only $\sim 3\%$ of the total,
having no significant effect to any kinematical distribution. Therefore,
these processes were not included in the {\sc pythia} calculation.

All calculations are subject to the experimental kinematic and isolation
cuts. {\sc diphox} accounts for the $gg\rightarrow\gamma\gamma$ process
in LO only. The predictions of {\sc resbos} are restricted to the invariant
mass range from $2m_{b}=9$ GeV/c$^{2}$ to $2m_{t}=350$ GeV/c$^{2}$, where
$m_{b}$ and $m_{t}$ are the masses of the bottom and top quarks, respectively.
NLO theoretical uncertainties are estimated by varying the fragmentation
(in {\sc diphox} only), renormalization and factorization scales up and down
by a factor of two relative to the default scale $\mu=M/2$ of {\sc diphox}
and $\mu=M$ of {\sc resbos}, and for the NLO PDF uncertainties (in both
{\sc diphox} and {\sc resbos}) by using the 20 CTEQ6.1M eigenvectors
\cite{cteq6m}. Theoretical uncertainties are shown only in the plots of
the relative deviations of the data from each calculation, in the form
(data$-$theory)/theory, as functions of the selected variable. No relative
deviations are shown for the {\sc pythia} $\gamma\gamma$ calculation. The
benchmark parton showering MC calculation, compared in detail with the data,
is {\sc pythia} $\gamma\gamma +\gamma{\rm j}$.

\section{Results}
The measured total cross section is shown in Table \ref{tab:tot-xsec} together
with the predictions from the three theoretical calculations. All three
calculations are consistent with the size of the measured cross section within
the experimental uncertainties.

%

\begin{table}[tbp]
\begin{center}
\caption{The total diphoton production cross section obtained from the
         measurement and from the theoretical calculations. The {\sc pythia}
         $\gamma\gamma$ calculation involves only the
         $q\bar q\rightarrow\gamma\gamma$ and $gg\rightarrow\gamma\gamma$
         processes. The {\sc pythia} $\gamma\gamma +\gamma{\rm j}$ calculation
         includes also the $q\bar q\rightarrow\gamma\gamma g$ and
         $gq\rightarrow\gamma\gamma q$ processes.}
\label{tab:tot-xsec}
\begin{tabular}{lc}
\hline\hline
 & Cross section (pb) \\ \hline
Data & $12.47\pm 0.21_{\rm stat}\pm 3.74_{\rm syst}$ \\
{\sc resbos} & $11.31\pm 2.45_{\rm syst}$ \\
{\sc diphox} & $10.58\pm 0.55_{\rm syst}$ \\
{\sc pythia} $\gamma\gamma +\gamma{\rm j}$ & $9.19$ \\
{\sc pythia} $\gamma\gamma$ & $5.03$ \\
\hline\hline
\end{tabular}
\end{center}
\end{table}

\begin{figure*}[t]
\includegraphics[width=7.5cm]{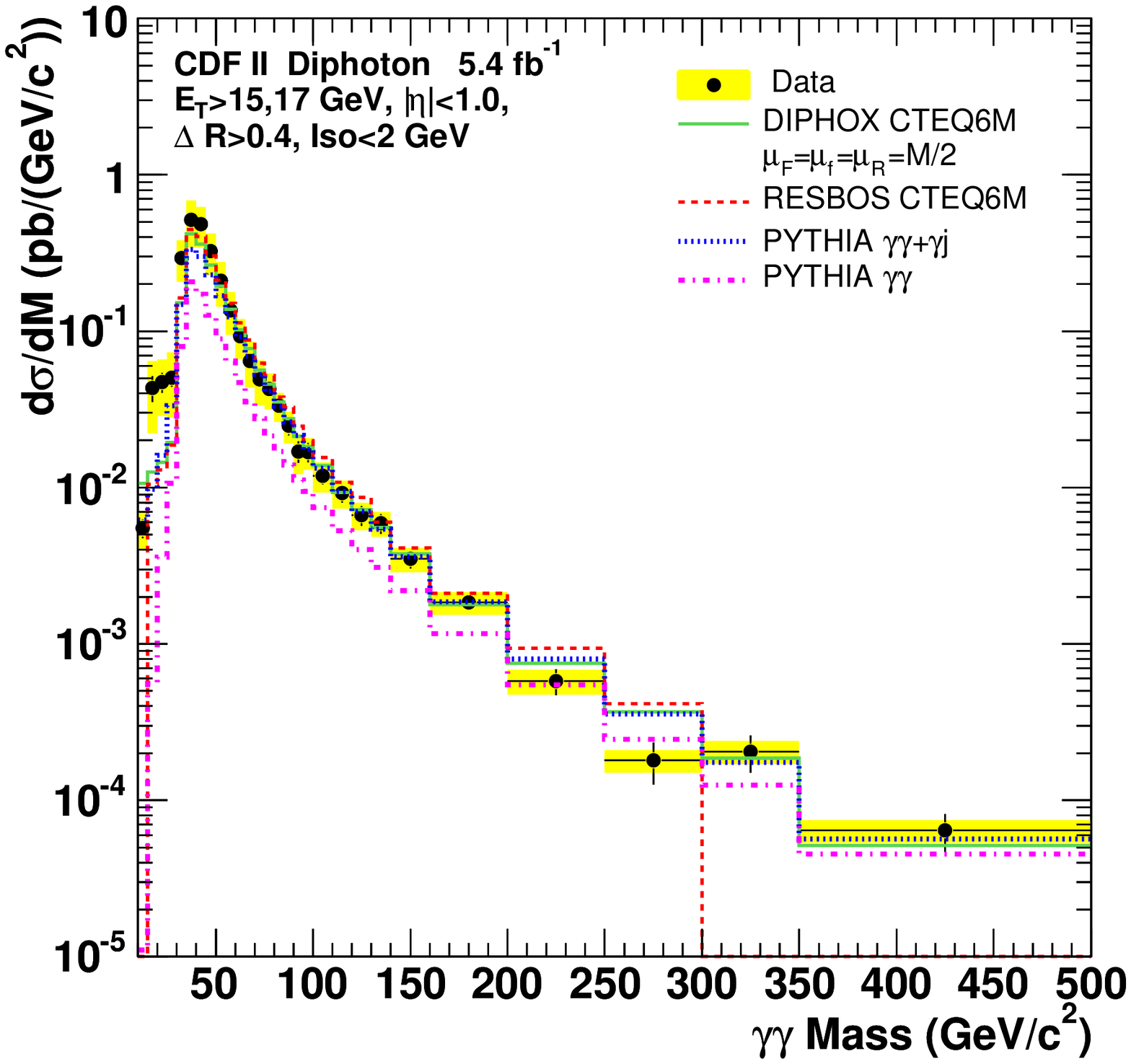}\hspace*{0cm}
\includegraphics[width=7cm]{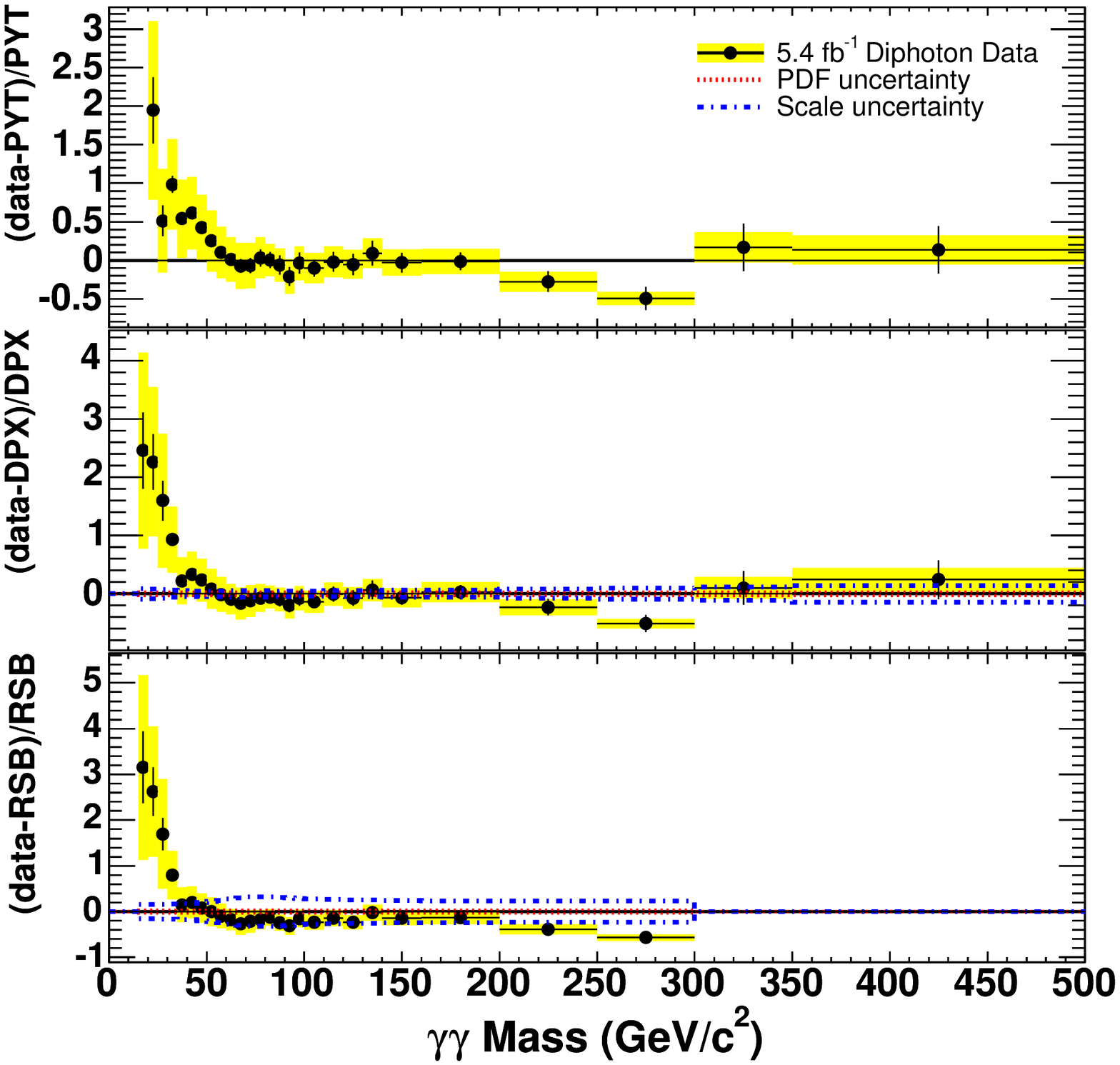}
\vspace*{0cm}
\includegraphics[width=7.5cm]{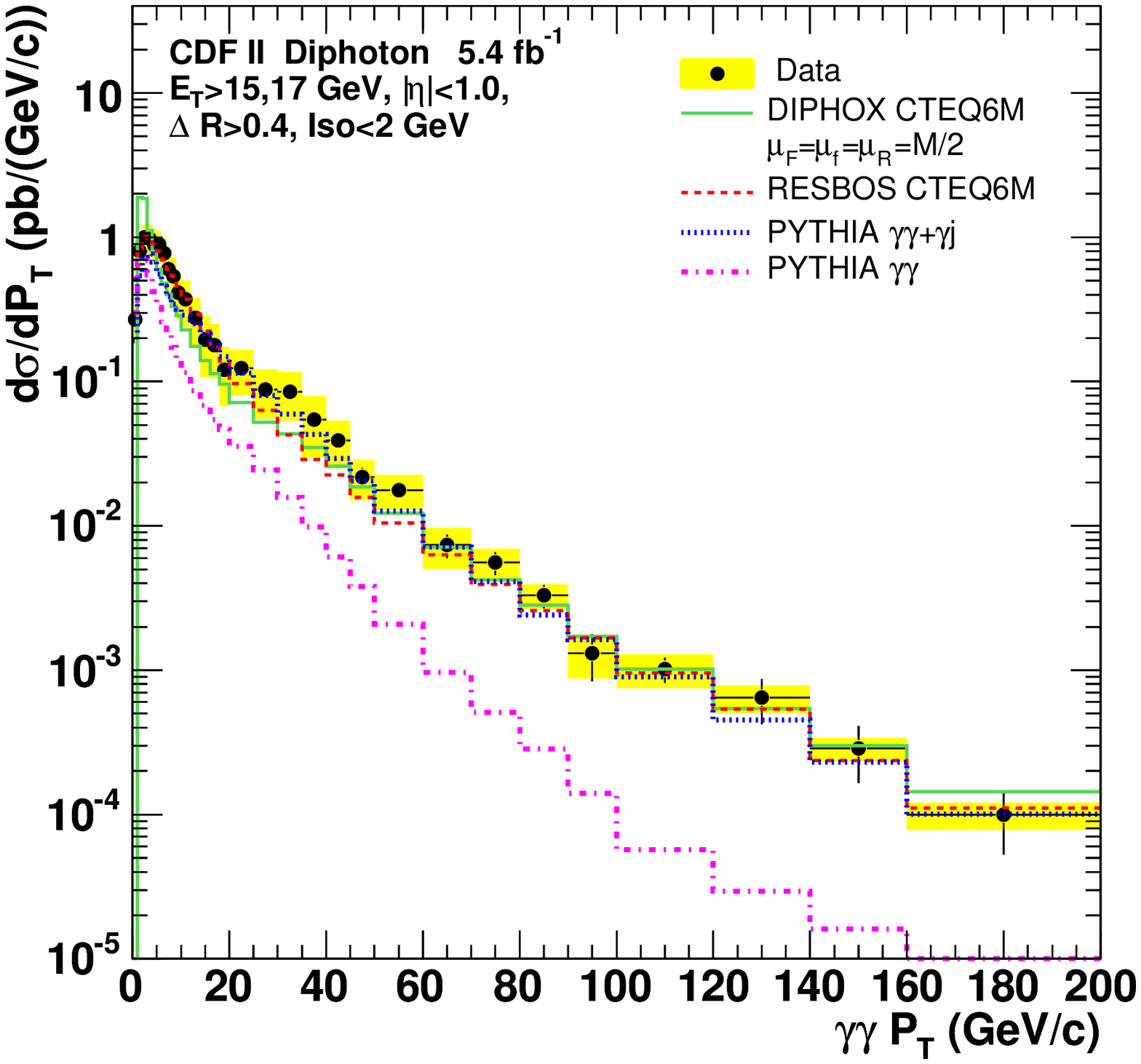}\hspace*{0cm}
\includegraphics[width=7cm]{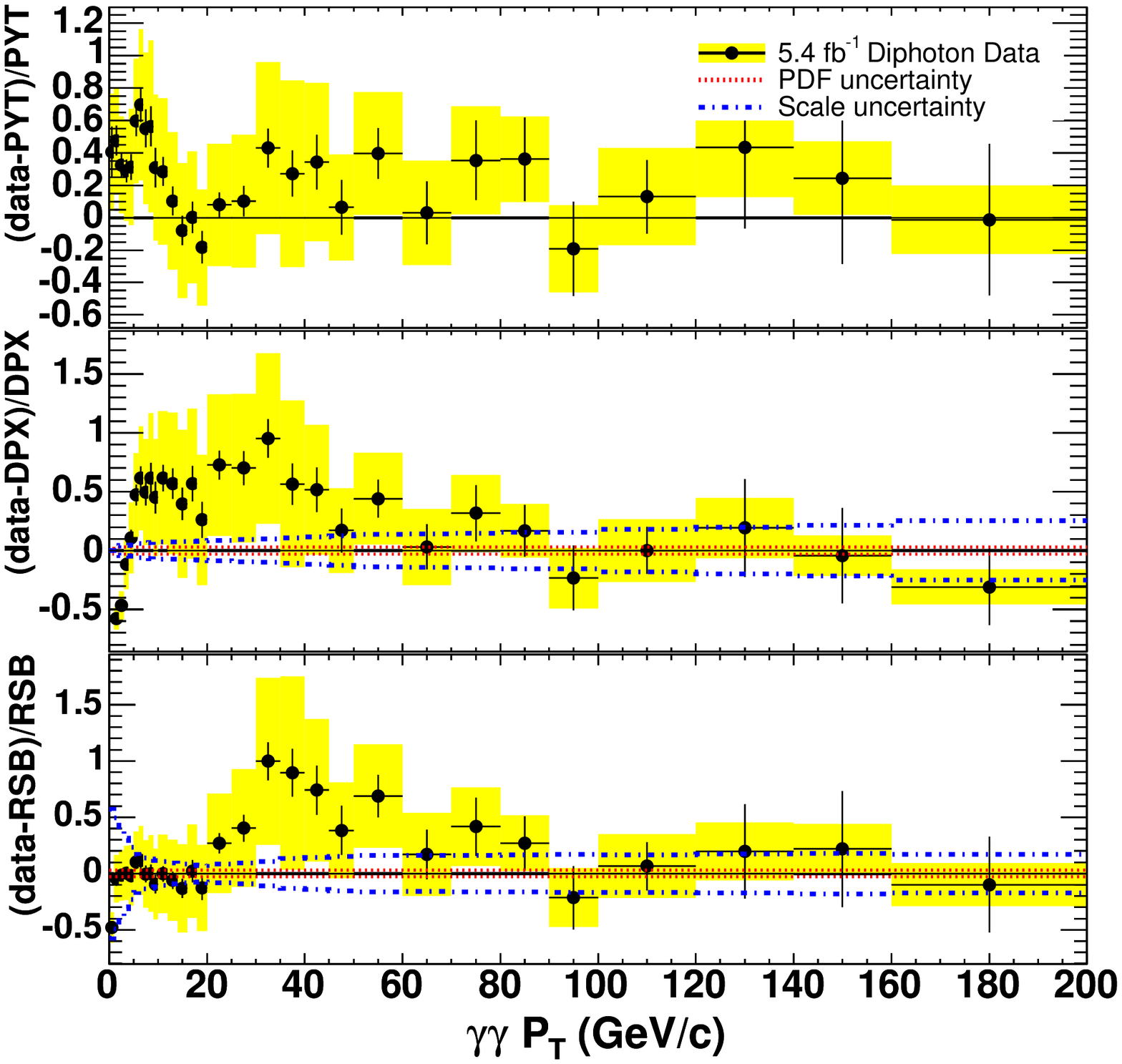}
\vspace*{0cm}
\includegraphics[width=7.5cm]{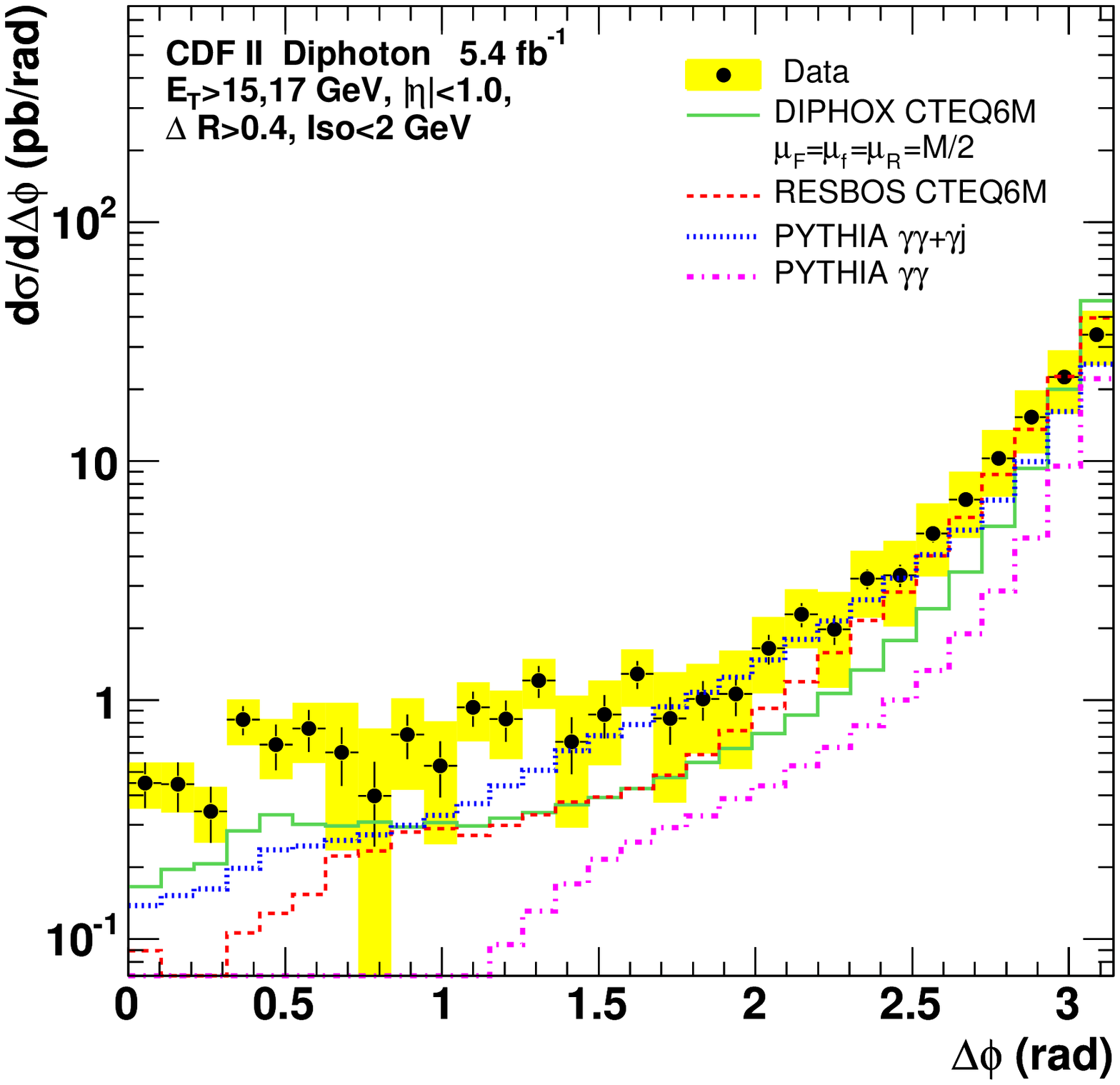}\hspace*{0cm}
\includegraphics[width=7cm]{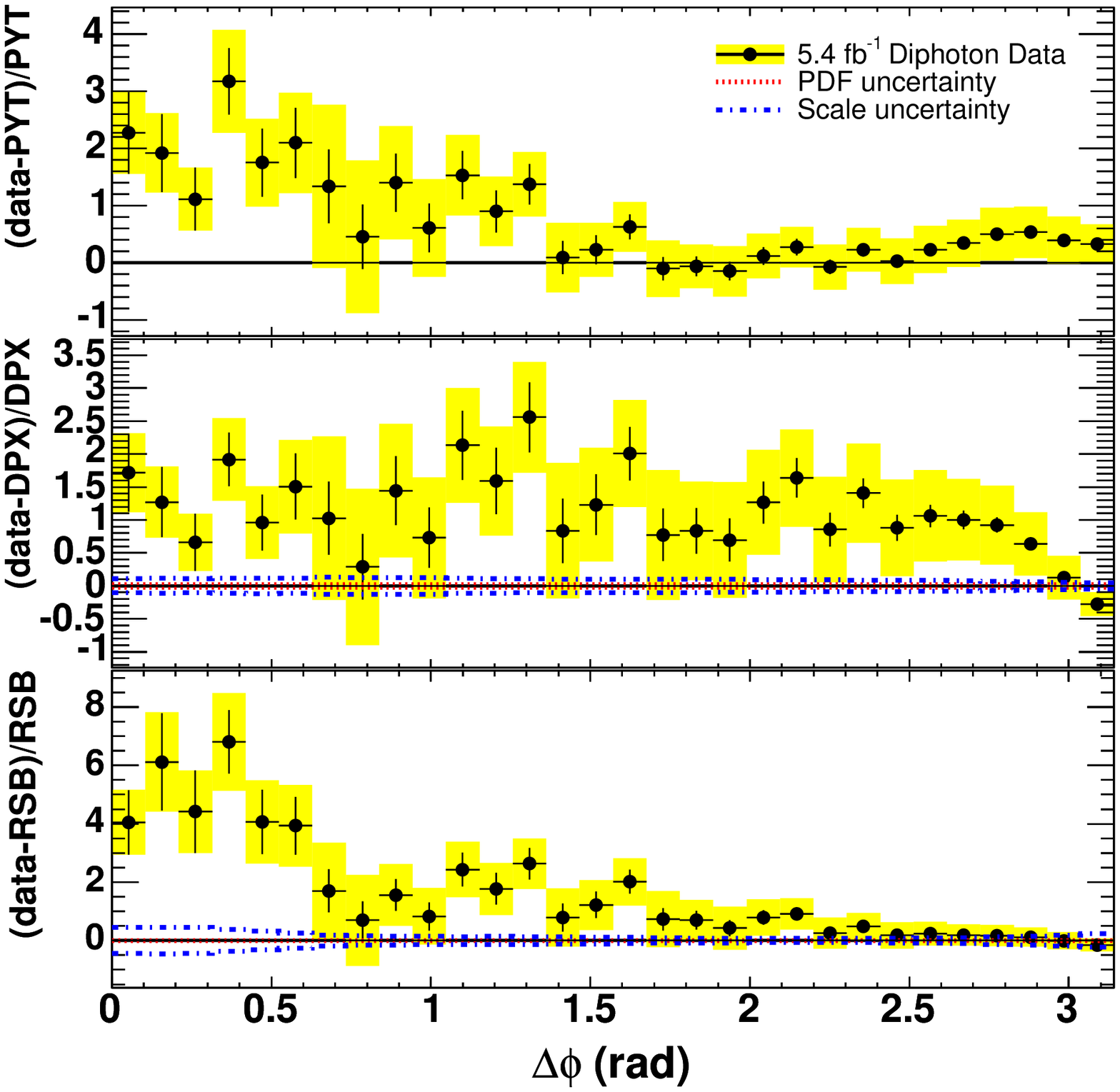}
\vspace*{-0.1cm}
\caption{The measured differential cross sections compared with three
theoretical predictions discussed in the text. The left windows show the
absolute comparisons and the right windows show the fractional deviations
of the data from the theoretical predictions. Fractional deviations for
{\sc pythia} refer to the $\gamma\gamma+\gamma{\rm j}$ calculation. Note
that the vertical axis scales differ between fractional deviation plots.
The comparisons are made as functions of the diphoton mass (top), transverse
momentum (middle) and azimuthal angle difference (bottom). The shaded area
around the data points indicates the total systematic uncertainty of the
measurement.\label{fig:results1}}
\end{figure*}

Fig. \ref{fig:results1} shows the comparison between the measured and
predicted diphoton cross sections as functions of the diphoton invariant mass
$M$, the diphoton transverse momentum $P_{\rm T}$ and the difference
$\Delta\phi$ between the azimuthal angles of the two photons in the event.
While the {\sc pythia} direct calculation ($\gamma\gamma$) fails to describe
both the scale and shape of the data, including radiation brings the prediction
in fair agreement with the data. In particular, radiation makes the $P_{\rm T}$
and $\Delta\phi$ distributions harder because of the presence of at least one
hard jet in the final state of events in which one photon originates from
radiation. The mass distributions show a reasonable agreement with the data
for all predictions above the peak at 30 GeV/c$^{2}$, particularly in the
region 80 GeV/c$^{2}$ $<M<$ 150 GeV/c$^{2}$ relevant to searches for the Higgs
boson \cite{higgs}. However, all predictions underestimate the data around and
below the peak. In the $P_{\rm T}$ spectrum all predictions underestimate the
data in the region between 20 and 50 GeV/c, a feature also observed in the
earlier measurements \cite{cdfdip,diphox}. For $P_{\rm T}$$<$20 GeV/c, where
soft gluon resummation is most important, only the {\sc resbos} prediction
describes the data. Discrepancies between data and theory are most prominent
in the comparison of the measured and predicted distributions of $\Delta\phi$.
In this case all three predictions fail to describe the data across the whole
spectrum. Approaching $\Delta\phi=\pi$, where soft gluon processes are expected
to manifest, the {\sc resbos} prediction agrees better with the data. In the
range 1.4 rad$<$$\Delta\phi$$<$2.2 rad only the {\sc pythia} prediction
describes the data and remains closest to the data down to 1 rad. In the
low $\Delta\phi$ tail, which corresponds to the region of low $M$
($<$50 GeV/c$^{2}$), all three predictions are lower than the data, although
the {\sc diphox} prediction, by explictly including non-perturbative
fragmentation, lies closer to the data for $\Delta\phi$$<$1 rad.

\begin{figure*}[t]
\includegraphics[width=7.5cm]{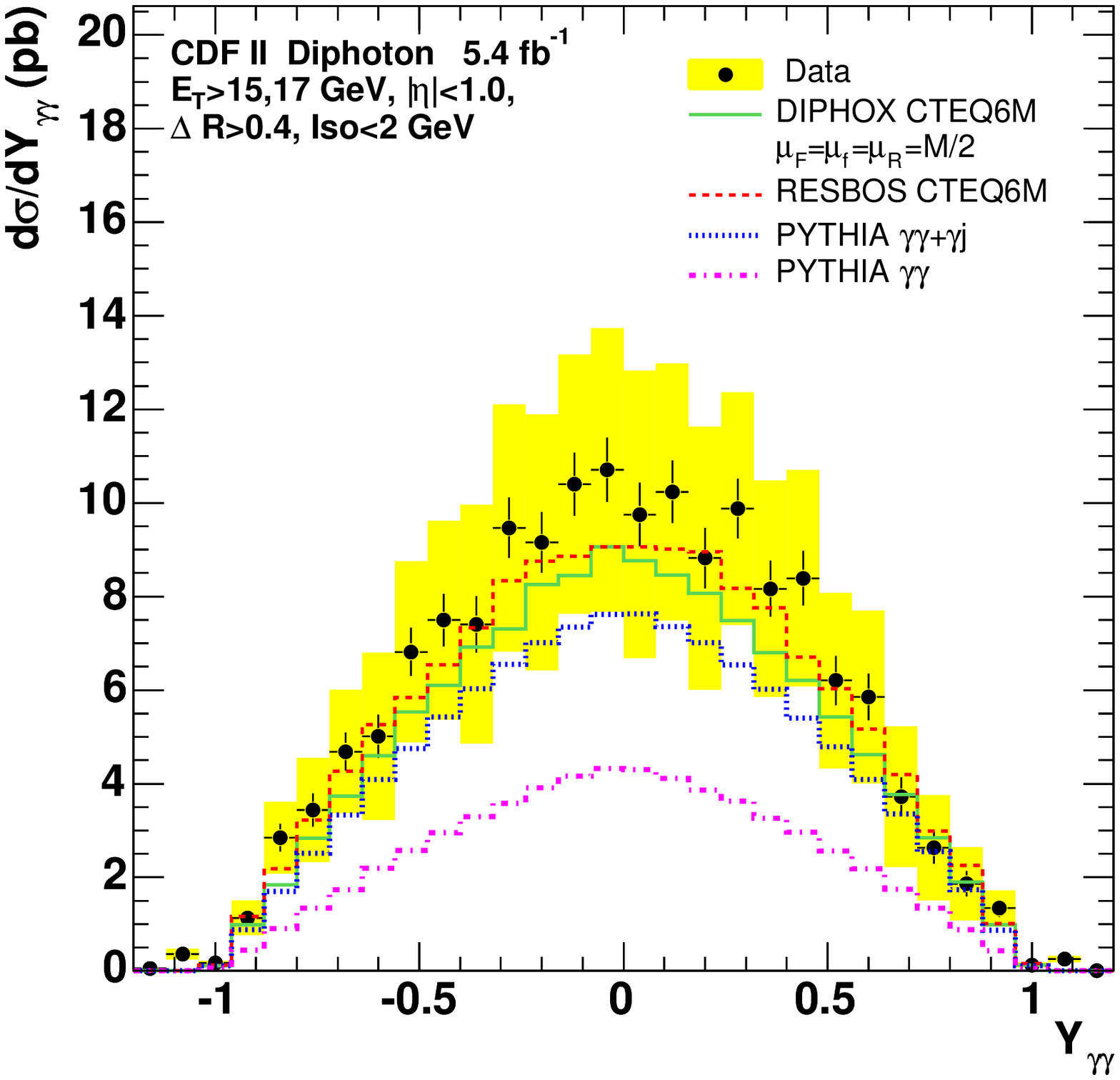}\hspace*{0cm}
\includegraphics[width=7cm]{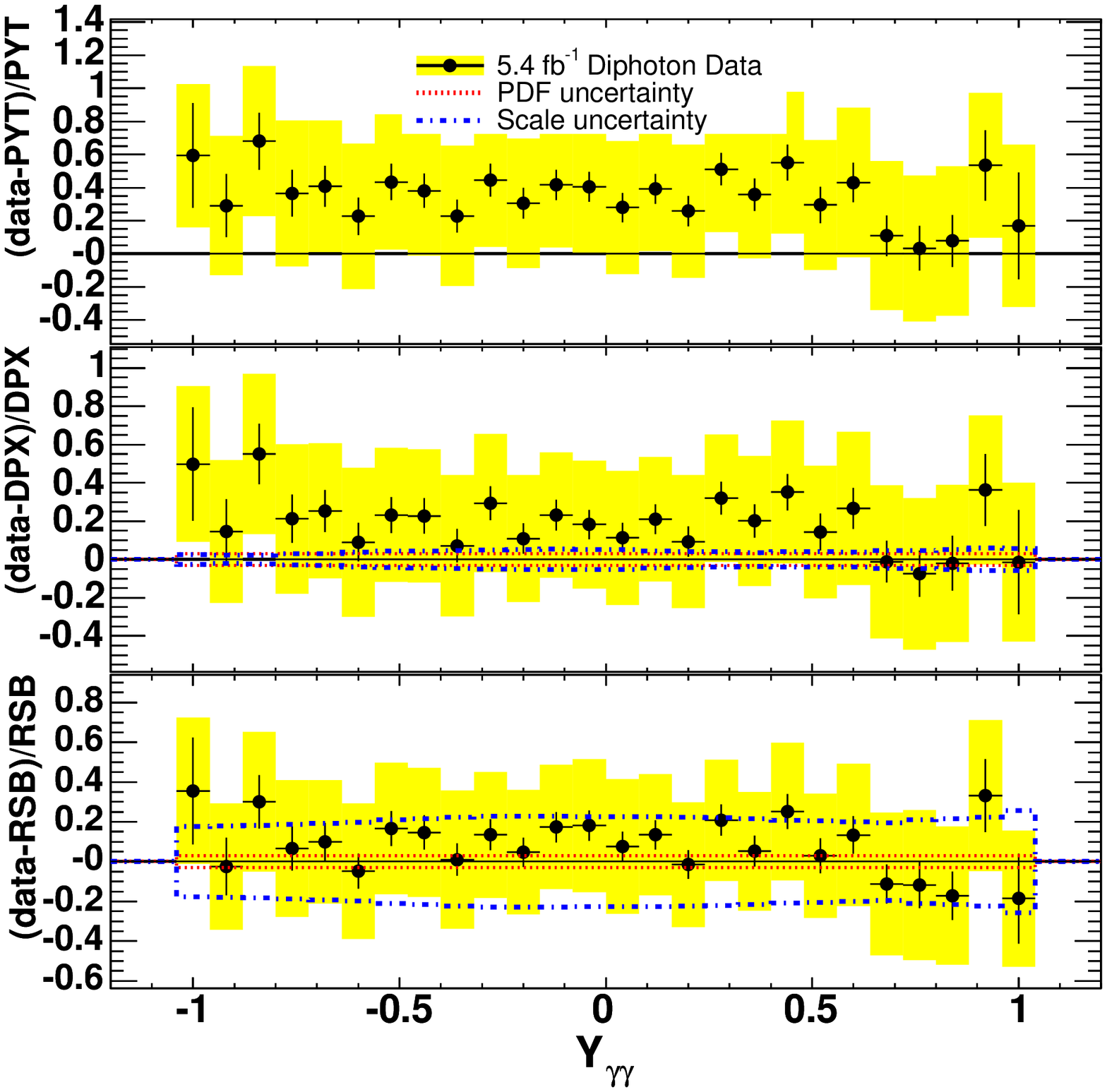}
\vspace*{0cm}
\includegraphics[width=7.5cm]{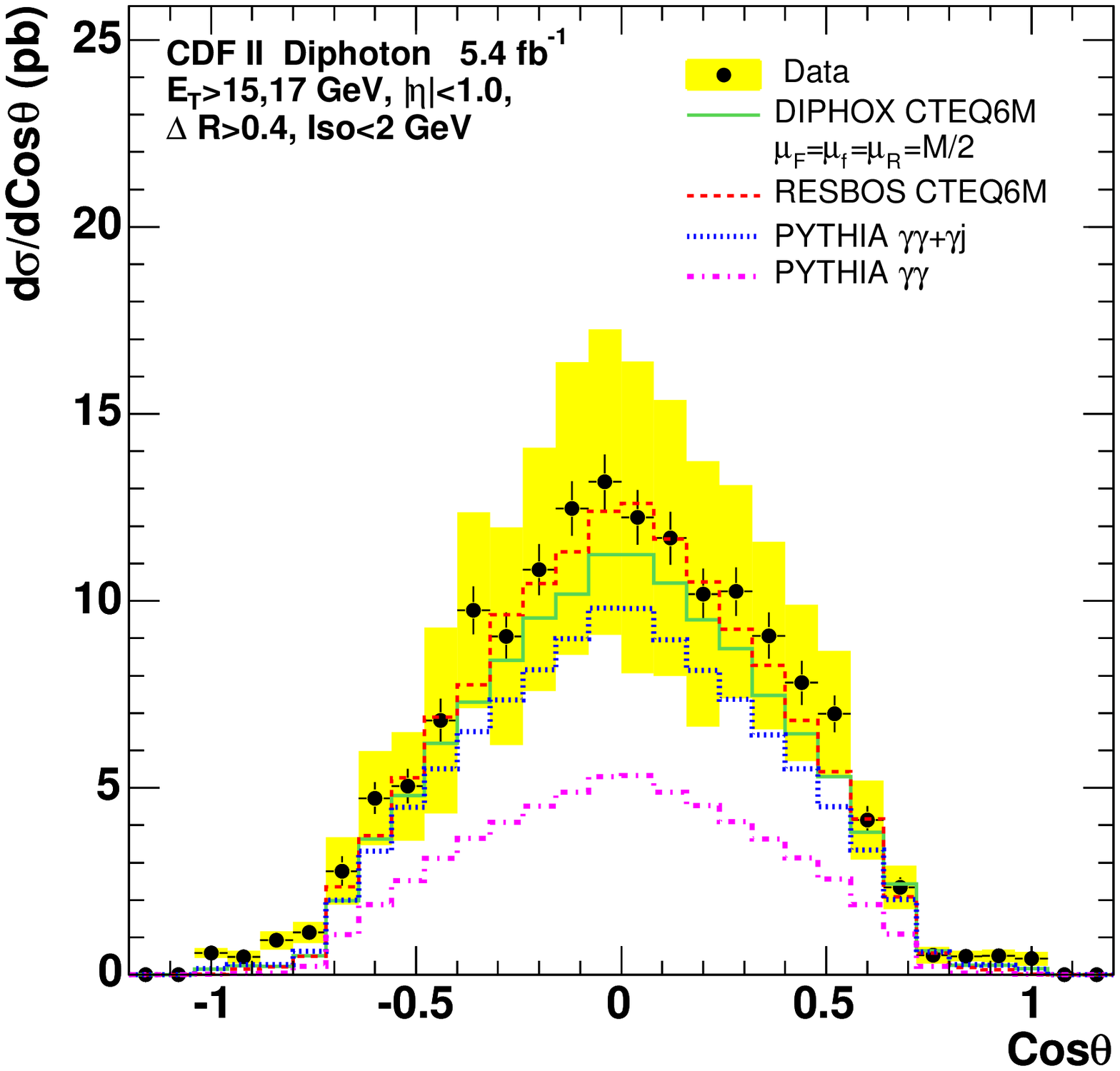}\hspace*{0cm}
\includegraphics[width=7cm]{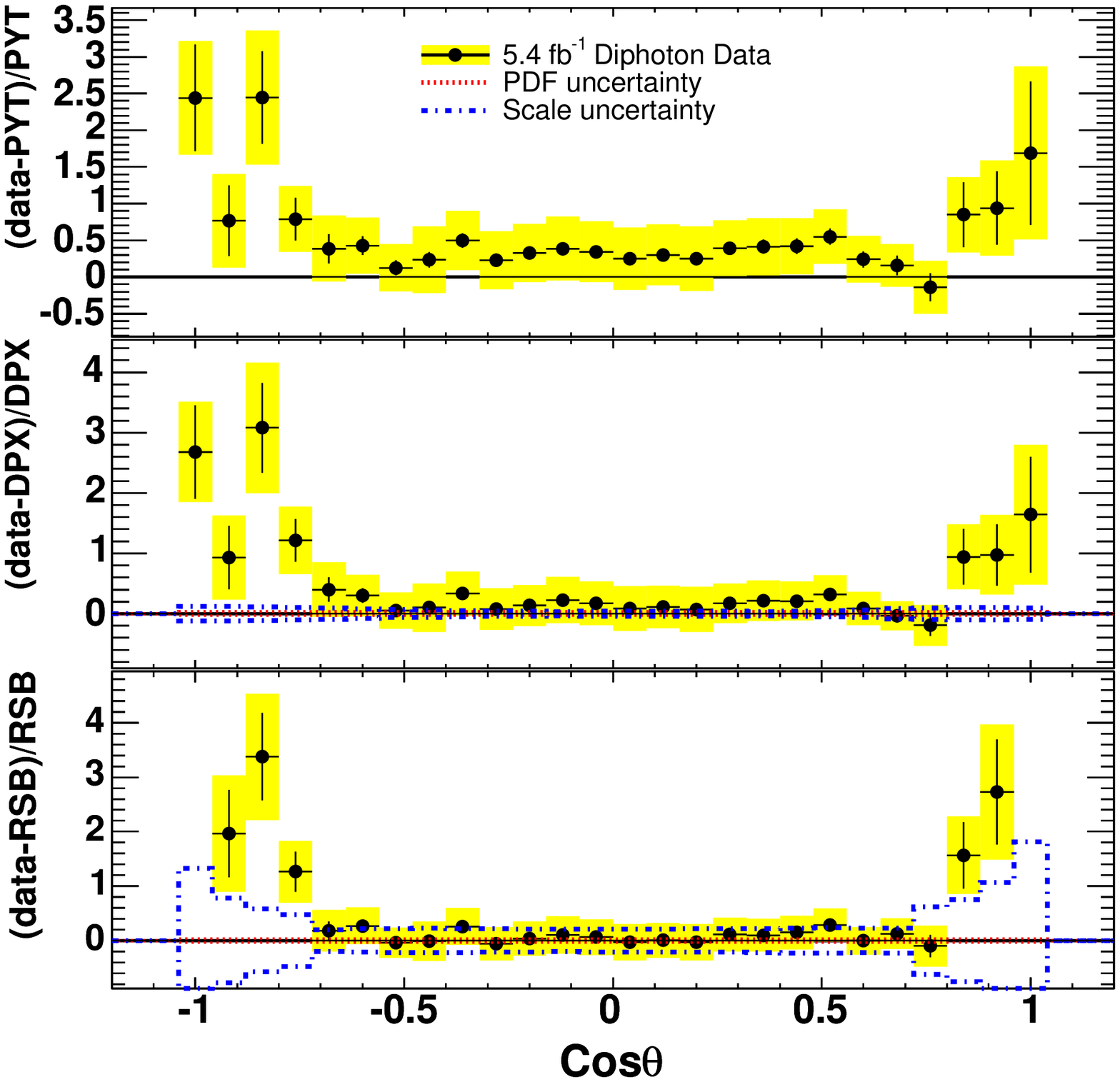}
\vspace*{0cm}
\includegraphics[width=7.5cm]{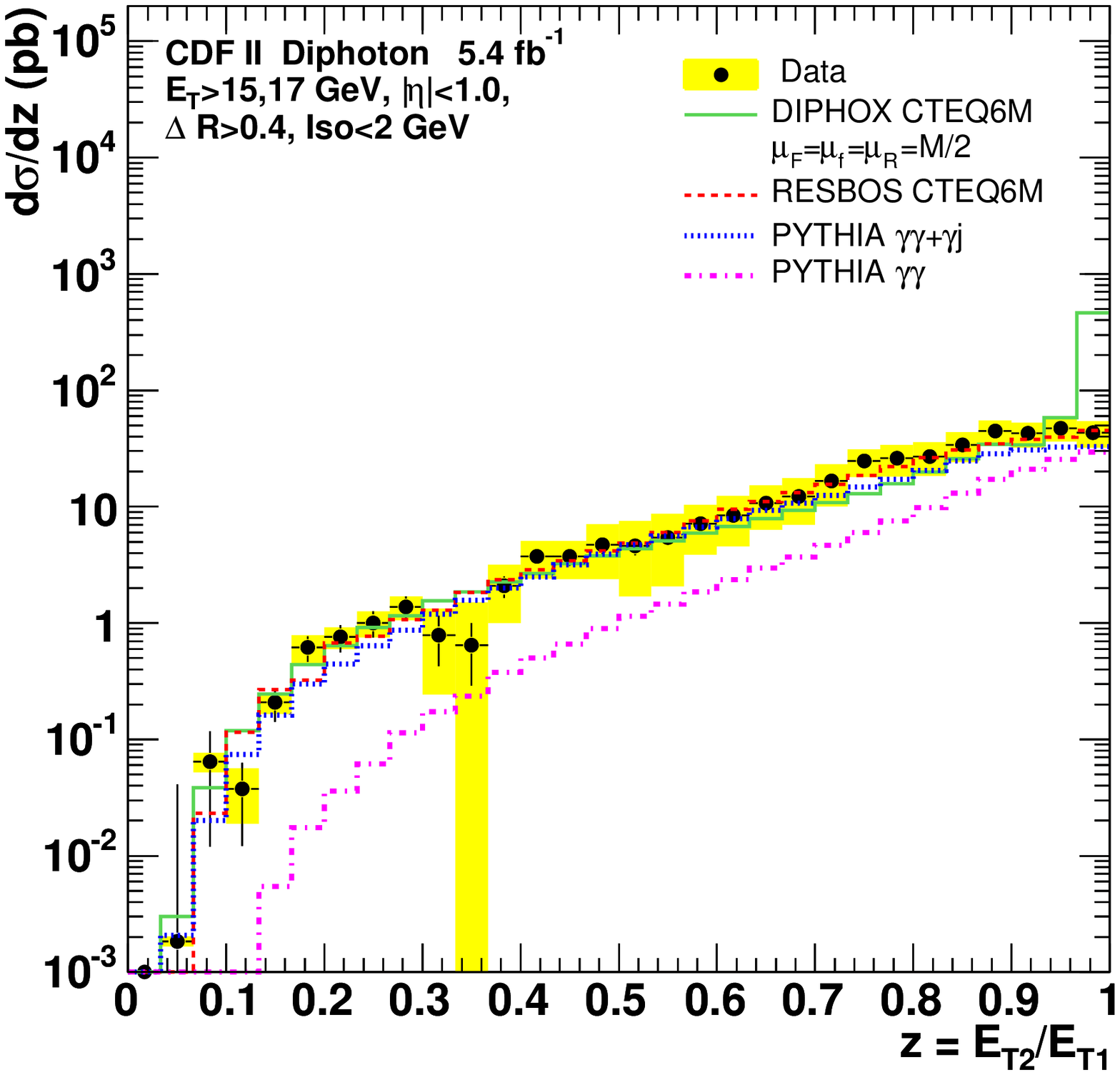}\hspace*{0cm}
\includegraphics[width=7cm]{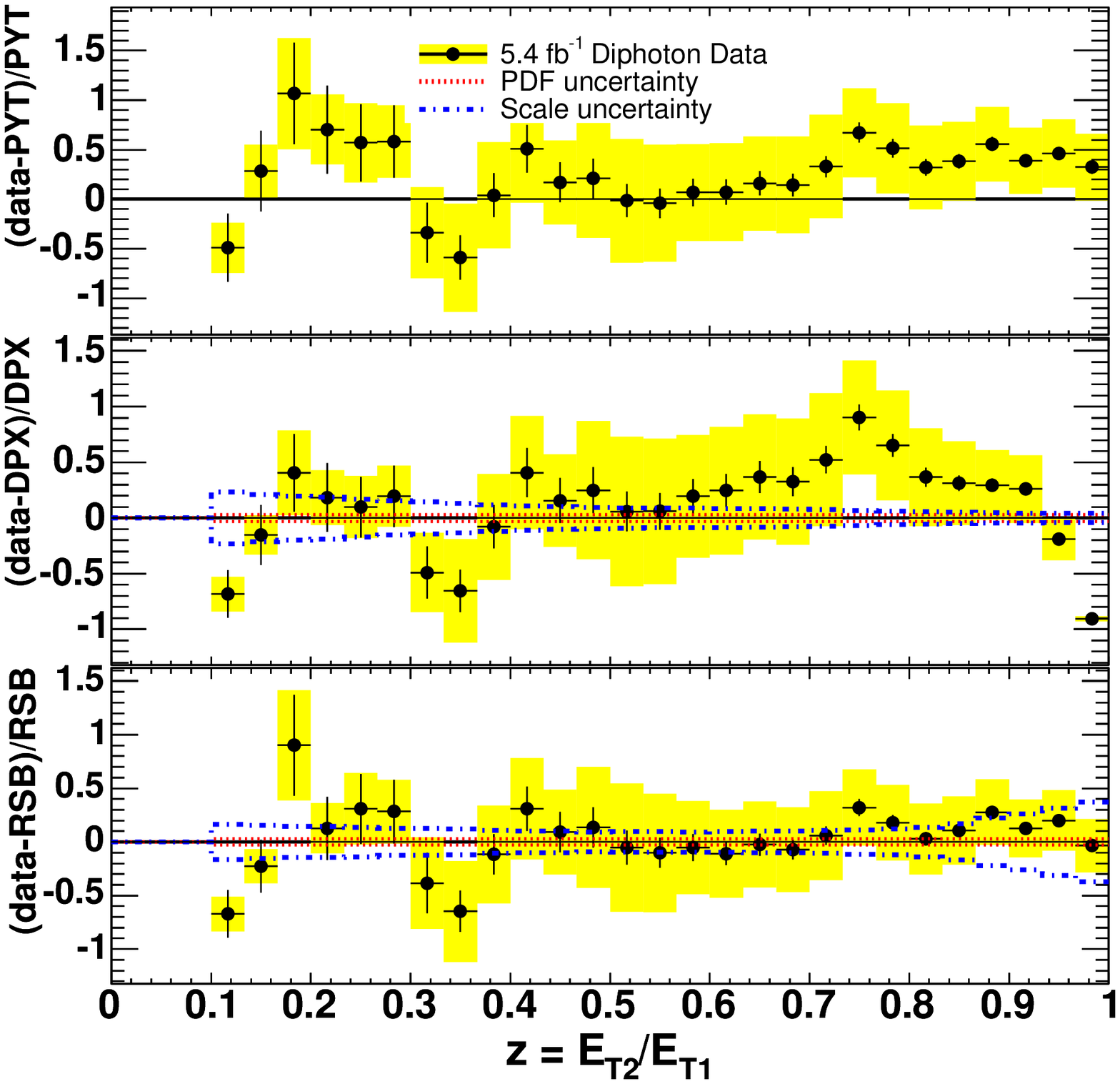}
\vspace*{-0.1cm}
\caption{The measured differential cross sections compared with three
theoretical predictions discussed in the text. The left windows show the
absolute comparisons and the right windows show the fractional deviations
of the data from the theoretical predictions. Fractional deviations for
{\sc pythia} refer to the $\gamma\gamma+\gamma{\rm j}$ calculation. Note
that the vertical axis scales differ between fractional deviation plots.
The comparisons are made as functions of the diphoton rapifity (top),
cosine of the polar angle in the Collins-Soper frame (middle) and ratio
of the subleading photon $E_{\rm T}$ to leading photon $E_{\rm T}$ (bottom).
The shaded area around the data points indicates the total systematic
uncertainty of the measurement.\label{fig:results2}}
\end{figure*}

Fig. \ref{fig:results2} shows the comparison between the measured and
predicted diphoton cross sections as functions of the diphoton rapidity
$Y_{\gamma\gamma}$, the cosine of the polar angle in the Collins-Soper frame
$\cos\theta$ \cite{CS}, and the ratio $z$ of the subleading photon $E_{\rm T}$
to leading photon $E_{\rm T}$. All three predictions agree fairly well with the
measured $d\sigma/dY_{\gamma\gamma}$, $d\sigma/d\cos\theta$ and $d\sigma/dz$,
within uncertainties. Exceptions are the predictions of all three calculations
underestimating the data in the two ends of the $\cos\theta$ spectrum, where
again gluon scattering processes and associated fragmentation are expected
to dominate \cite{resbos}.

In general, all three calculations reproduce most of the main features of
the data, as observed in the earlier diphoton cross section measurements
\cite{cdfdip,d0}. However, depending on their approximations, they display
differences with each other and with the data in certain kinematic regions.
There is a problem common to all three calculations in the description of
events with very low diphoton mass, low azimuthal distance and diphoton
transverse momentum in the region of the so-called ``Guillet shoulder''
(20 GeV/$c$$<$$P_{\rm T}$$<$50 GeV/$c$) \cite{diphox}. Such events include
fragmentation at a relatively high rate. The {\sc pythia} $\gamma\gamma$
calculation fails completely to describe the data both in the scale, where
it is low by a factor of 2.5 (see Table \ref{tab:tot-xsec}), and in the shape,
particularly of the $P_{\rm T}$, $\Delta\phi$ and $z$ distributions, where it
predicts a much softer spectrum than the data. This is in agreement with the
conclusion of Ref. \cite{cdfdip} which tested only {\sc pythia} $\gamma\gamma$
as a parton showering MC prediction.

\section{Summary}
In summary, the diphoton production cross section, differential in
kinematic variables sensitive to the reaction mechanism, is measured
using data corresponding to an integrated luminosity of 5.36 fb$^{-1}$
collected with the CDF II detector. The high statistics of the measured
sample allows for a higher precision scan over a much more extended phase
space than previous measurements. The overall systematic uncertainty is
limited to about 30\%. The results of the measurement are compared with
three state-of-the-art calculations, applying complementary techniques in
describing the reaction. All three calculations, within their known
limitations, reproduce the main features of the data, but none of them
describes all aspects of the data. The inclusion of photon radiation in the
initial and final states significantly improves the {\sc pythia} parton
shower calculation (see the left-hand panels in Figures \ref{fig:results1}
and \ref{fig:results2}), which is suitable for background simulations in
searches for a low mass Higgs boson and new phenomena.

\begin{acknowledgments}
We thank the Fermilab staff and the technical staffs of the participating
institutions for their vital contributions. We also thank P. Nadolsky, C.-P.
Yuan, Z. Li, J.-P. Guillet, C. Schmidt and S. Mrenna for their valuable help
in the theoretical calculations. This work was supported by the U.S.
Department of Energy and National Science Foundation; the Italian Istituto
Nazionale di Fisica Nucleare; the Ministry of Education, Culture, Sports,
Science and Technology of Japan; the Natural Sciences and Engineering Research
Council of Canada; the National Science Council of the Republic of China; the
Swiss National Science Foundation; the A.P. Sloan Foundation; the
Bundesministerium f\"ur Bildung und Forschung, Germany; the Korean World Class
University Program, the National Research Foundation of Korea; the Science and
Technology Facilities Council and the Royal Society, UK; the Institut National
de Physique Nucleaire et Physique des Particules/CNRS; the Russian Foundation
for Basic Research; the Ministerio de Ciencia e Innovaci\'{o}n, and Programa
Consolider-Ingenio 2010, Spain; the Slovak R\&D Agency; the Academy of
Finland; and the Australian Research Council (ARC). 
\end{acknowledgments}

\bigskip 

\begin{thebibliography}{9}
\bibitem{caliso}
The calorimeter isolation is defined as the difference of the transverse
energy $E_{\rm T}=E\sin\theta$ in the isolation cone minus the transverse
energy in the tower cluster of the photon. The isolation cone is defined
to have a radius $R=\sqrt{(\Delta\eta)^{2}+(\Delta\phi)^{2}}=0.4$ around
the axis of the shower profile, where $\eta=-\ln[\tan(\theta/2)]$ is the
pseudo-rapidity, $\theta$ is the polar angle and $\phi$ the azimuth in the
coordinate system of the laboratory with polar axis along the colliding
beams and origin at the center of the detector. The isolation cut implies
that the angular separation of the two photons is $\Delta R\ge 0.4$.
\bibitem{trkiso}
The track isolation is defined as the scalar sum of the transverse momenta
$p_{\rm T}=\vert\vec{p}\vert\sin\theta$ of all tracks originating from the
primary vertex of the event and lying within the photon reconstruction cone.
\bibitem{cones}
The complementary cone is defined by rotating the photon reconstruction
cone by an azimuthal angle of 90$^{\circ}$ about the beam axis, keeping
the polar angle fixed.
\end{thebibliography}

\end{document}